# Mapping of the system of software-related emissions and shared responsibilities

Laura Partanen
*dept. of Software Engineering*
*LUT University*
Lahti, Finland
laura.partanen@lut.fi

Antti Sipilä
*dept. of Responsibility*
*TIEKE*
Helsinki, Finland
antti.sipila@tieke.fi

Md Sanaul Haque
*dept. of Software Engineering*
*LUT University*
Lappeenranta, Finland
sanaul.haque@lut.fi

Jari Porras
*dept. of Software Engineering*
*LUT University / Aalto University / University of Huddersfield*
Lappeenranta, Finland / Helsinki, Finland / Huddersfield, UK
jari.porras@lut.fi

*Abstract*— The global climate is experiencing a rapid and unprecedented warming trend. The ICT sector is a notable contributor to global greenhouse gas emissions, with its environmental impact continuing to expand. Addressing this issue is vital for achieving the objectives of the Paris Agreement, particularly the goal of limiting global temperature rise to 1.5°C. At the European Union level, regulatory measures such as the CSRD and the CSDD impose obligations on companies, including those within the ICT sector, to recognize and mitigate their environmental footprint. This study provides a comprehensive system mapping aimed at enhancing the awareness and understanding of software-related emissions and the corresponding responsibilities borne by the ICT sector. The mapping identifies the primary sources of carbon emissions and energy consumption within the ICT domain while also outlining the key responsibilities of the stakeholders accountable throughout the software lifecycle.

## I. Introduction

According to the Intergovernmental Panel on Climate Change (IPCC)[1], greenhouse gases from human activities have increased the global temperature by 1.1°C since 1850-1900. With the Paris Agreement[2], nations worldwide have committed to *"hold the increase in the global average temperature to well below 2°C above pre-industrial levels´ and pursue efforts ´to limit the temperature increase to 1.5°C above pre-industrial levels´."* To achieve this 1.5°C target, emissions need to reach net-zero by 2050 [1]. In Europe, this commitment is emphasized through the European Green Deal[3], which aims for carbon neutrality by 2050. This objective is regulated by the European Climate Law[4]. In Finland, the goal for carbon neutrality is set to 2035 [2].

The ICT sector is estimated to account for 2.1–3.9% of global greenhouse gas emissions [3]. While it is acknowledged that the Information and Communication Technology (ICT) sector has significant potential to reduce carbon emissions from other sectors, known as handprint effects, its own carbon footprint needs attention [4]. According to the GeSI SMARTer2030 report [5], the ICT's carbon footprint is the fastest growing among all industries and is projected to triple between 2015 and 2025.

The Finnish Ministry of Transport and Communications published the Climate and Environmental Strategy for the ICT Sector in March 2021 as the first country in the world [6]. This pioneering strategy addresses both carbon handprint and footprint effects in six different areas of ICT (Hardware, Software, Resources, Measuring, Awareness, and Emerging technologies), providing propositions for corresponding stakeholder actions. However, it's worth noting that the strategy does not explicitly consider the responsible stakeholders for the emissions.

This study aims to understand the system of software-related emissions and to investigate the relationships between the main emission sources and the pertinent stakeholders. Additionally, the study seeks to provide an analysis of the responsibilities assigned to these stakeholders. Consequently, the following research question emerges:

- What are the primary sources of software-related emissions, how are these sources interconnected, and what type of responsibilities do the relevant stakeholders have in managing these emissions?

In addition to addressing global warming and climate change, it is essential to also prioritize attention to other environmental challenges, such as biodiversity loss or water consumption. The authors of this paper want to emphasize the acknowledgment of this, even though this paper concentrates on the emissions and energy consumption of the ICT sector. There is rather new research on environmental impacts with environmental footprint calculation, where climate change and $CO_2$ emissions play one role among land and sea usage, direct exploitation of organisms, pollution, and invasive alien species [7].

The rest of this paper is structured as follows. Section II presents the research context of the study and provides the theoretical background on which the research builds. Section III describes the research process, followed by the results of the systems mapping in section IV. Section V discusses the types of responsibilities in the mapping. Finally, section VI concludes this paper.

## II. Background

This background section introduces the research context – the regulation behind the sustainability issues and two Finnish projects in which the research question was raised. Section C-E introduces the theoretical context of the study. The system mapping presented later in the paper in Section IV builds on this theoretical background.

---

[1] https://www.ipcc.ch/2021/08/09/ar6-wg1-20210809-pr/
[2] https://unfccc.int/process-and-meetings/the-paris-agreement
[3] https://commission.europa.eu/strategy-and-policy/priorities-2019-2024/european-green-deal_en
[4] https://climate.ec.europa.eu/eu-action/european-climate-law_en

*A. Regulation*

Since 2014, the EU's Non-Financial Reporting Directive (NFRD) [5] mandates that large public interest entities with more than 500 employees – such as banks, insurance firms, and major listed companies – submit a non-financial statement. This statement must provide sufficient information to assess the company's development, performance, position, and the impact of its activities, specifically addressing environmental, social, and employee issues, as well as matters concerning human rights, anti-corruption, and bribery. The Corporate Sustainability Reporting Directive (CSRD) [6], which came into effect on January 2023, expands the scope of mandatory reporting. It enhances and updates the requirements for companies to disclose social and environmental information, extending these obligations to a broader range of large companies and listed small and medium-sized enterprises (SMEs). The financial year 2024 is the first year to report under the CSRD obligations. Companies are obligated to report direct (scope I) and indirect (scope II & III) emissions. Software-related emissions are included in Scope III [8].

Since the proposal of the Corporate Sustainability Due Diligence Directive [7] (CSDD) is proceeding after the agreement in March 2024 of the Council of the European Union, in the future companies will be obligated to identify and prevent negative environmental impacts alongside human rights. Also with the proposal of Green Claims Directive [8] aiming to prevent greenwashing, there is obvious pressure from the European Union on organizations to take sustainability into account in their operations.

*B. Finnish ICT projects*

The Climate and Environmental Strategy for the ICT Sector in Finland has been implemented e.g. via two Green ICT projects. The authors of this paper managed and actively participated in these projects in 2021-2023. Brief descriptions of these projects follow in the next sub-chapters.

*a) Project A*

The Green Metrics for Public Digitalization Acquisitions – MitViDi [9] project was executed from January 2022 to September 2023. The project aimed to develop a set of metrics for assessing the climate and environmental impacts of individual software, with a specific emphasis on public procurement. Consequently, a set of green requirements was published to be taken into the software procurement processes.

*b) Project B*

Green ICT ecosystem – building sustainability and competitiveness for businesses in the Uusimaa region [10] started in June 2021 and aimed to facilitate a collective transition toward sustainability by employing awareness-building, educational endeavors, and knowledge dissemination among different stakeholders. To accomplish its objectives, the project offered webinars, online workshops, and published guides for both procurers and producers. Additionally, a web-based self-assessment tool was developed for organizations to evaluate their carbon-neutral actions and establish a foundation for their development plans. The project concluded in August 2023.

*C. Sustainability*

The modern definition of sustainability builds on the Brundtland Report "Our Common Future" published in 1987 by the United Nations [9]. Sustainable development has been defined in the report as "development that meets the needs of the present without compromising the ability of future generations to meet their own needs." This approach emphasizes the balance between short-term human needs and long-term sustainability of environmental, social, and economic systems, aiming to prevent actions that could deplete resources, damage ecosystems, or create lasting inequalities. There are various frameworks presented after the Brundtland report, e.g. Circles of Sustainability [10] dividing sustainability into four dimensions, ecology, economics, politics, and culture. It is recognized that human actions have already crossed six of the nine planetary boundaries [11]. The pioneering Dasgupta Review highlights the need to understand the connection between the three sustainability pillars – all species, including humans, are dependent on nature [12]. Nature provides us the resources for everything we do, in business also.

Concerning the ICT or IT sector, sustainability has been conceptualized through various frameworks, including "Green IT", which addresses the environmental impact of the IT sector itself, and "Green by IT" which focuses on using IT to reduce the environmental footprint of other sectors [13]. Other terminologies, such as "Digital Sustainability" and "ICT for Sustainability (ICT4S)" encompass both environmental concerns and broader dimensions of sustainability, including social considerations [14]. Penzenstadler et al. [15] and the Karlskrona Manifesto [16] later expand the traditional three pillars of sustainability – environmental, social, and economic – by introducing two additional dimensions: technical, which pertains to the long-term viability of software, and individual, addressing personal autonomy and fulfillment. While sustainability may be categorized into three, four, or five dimensions, it is essential to recognize and understand the interrelationships and interdependencies among these dimensions as they collectively form a unified and integrated system. The Sustainability Awareness Framework (SusAF) [17] emphasizes this, providing a question-based framework for raising awareness of software sustainability within the five dimensions presented above.

*D. Software-related energy consumption and $CO_2$ emissions*

Three major sources for ICT sector emissions can be pointed out (Fig.1) – devices, data centers, and networks [5, 18, 19]. These emissions are formed from energy consumption. Projects A and B revealed a notable insight that

---

[5] https://eur-lex.europa.eu/legal-content/EN/TXT/?uri=celex%3A32014L0095

[6] https://eur-lex.europa.eu/legal-content/EN/TXT/?uri=CELEX%3A32022L2464

[7] https://eur-lex.europa.eu/legal-content/EN/TXT/?uri=CELEX:52022PC0071

[8] https://environment.ec.europa.eu/topics/circular-economy/green-claims_en

[9] https://tieke.fi/en/projects/green-metrics-for-public-digitalization-acquisitions-mitvidi/

[10] https://tieke.fi/en/projects/green-ict-project/

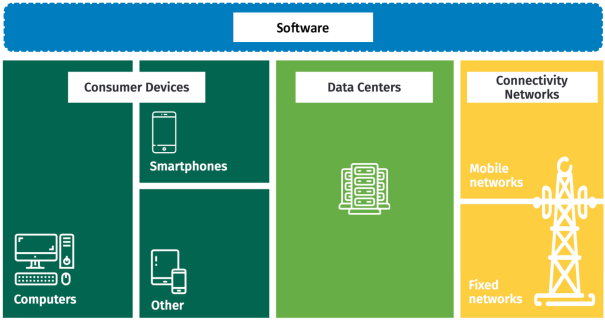

Fig. 1. Illustration of sources of ICT emissions according to The World Bank and ITU [19]. Software is added to the picture by the authors of this paper, not as an individual source of emissions but as a demanding component for the infrastructure.

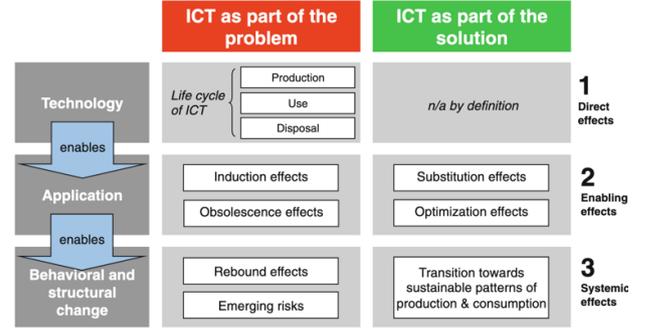

Fig. 2. The matrix of three-level effects of ICT according to Hilty & Aebischer [14].

TABLE I. ESTIMATES OF GLOBAL EMISSIONS OF THE ICT SECTOR [19].

| Industry | Emissions 2022/2020 (million tCO$_2$e) | | | Change 2022/2020 % | Electricity (TWh) | | | Change 2022/2020 % |
|---|---|---|---|---|---|---|---|---|
| | 2020 | 2021 | 2022 | | 2020 | 2021 | 2022 | 22/20 % |
| Telecommunications operators | 135 | 134 | 133 | -1% | 239 | 255 | 258 | 8% |
| Colocation data centers | 36 | 40 | 43 | 20% | 89 | 100 | 109 | 22% |
| Cloud & content | 22 | 27 | 32 | 46% | 54 | 70 | 85 | 63% |
| Subtotal | 193 | 201 | 208 | 8% | 382 | 425 | 442 | 18% |
| % of world | 0.6% | 0.6% | 0.6% | | 1.60% | 1.70% | 1.70% | |
| ICT Equipment | 154 | 173 | 154 | 0.5% | 282 | 329 | 311 | 10.6% |
| - PCs | 62 | 71 | 65 | 4.8% | 110 | 133 | 124 | |
| - Smartphones | 60 | 64 | 57 | -5.1% | 116 | 131 | 119 | 2.5% |
| - Network | 32 | 38 | 33 | 2.4% | 56 | 65 | 69 | 22.0% |
| Product use | 222 | 215 | 205 | -7.5% | 430 | 442 | 430 | -0.1% |
| - PCs | 203 | 197 | 187 | -7.9% | 394 | 405 | 392 | -0.5% |
| - Smartphones | 19 | 18 | 18 | -3.4% | 36 | 37 | 38 | 4.3% |
| Subtotal | 375 | 388 | 359 | -4.2% | 712 | 771 | 741 | 4.1% |
| % of world | 1.2% | 1.1% | 1.0% | | 3.0% | 3.0% | | |
| TOTAL | 568 | 589 | 567 | -0.2% | 1094 | 1196 | 1183 | 8.2% |
| % of world | 1.8% | 1.7% | 1.7% | | 4.6% | 4.7% | | |

most emissions come from the usage phase of the software life cycle [20], which can also be seen in Table I.

Table I illustrates the progress of impacts of colocation data centers, as well as the rapid growth in emissions and energy consumption of cloud and content data centers. Based on the information in Table I, one could say that it seems that other sources have somehow stabilized their level of emissions as well as the energy consumption in the years 2020–2022, excluding the energy consumption of the network. It can be stated that the core of these growing sources is data that is used by or collected through software.

The relation between the emission sources and software is straightforward – software requires all of the three (Fig. 1). Relation between software and the infrastructure can also be looked at from the opposite perspective, e.g., a smartphone is only an object made of several materials without software – to be usable, smartphones need software.

Even though this paper focuses on emissions and energy consumption, which are direct (first-order) effects, it is worth noticing that ICT and software also have enabling (second-order) effects and systemic (third-order) effects (see Fig. 2) [14]. This is important to be aware of when considering the overall effects of the ICT sector [17]. Using an e-commerce platform as an illustrative example, the three-level effects can be categorized as follows:

- First-order effects: the direct emissions and energy consumption associated with the data centers required to operate the platform. This includes powering servers, maintaining network infrastructure, and cooling systems.
- Second-order effects: indirectly, the platform enables increased consumer demand, contributing to higher emissions from manufacturing products and the transportation required for their delivery.
- Third-order effects: over time, the convenience of rapid delivery services alters consumer behavior and societal expectations, fostering a system that prioritizes speed, which can amplify environmental impacts through infrastructure expansion and resource-intensive logistics.

To estimate the carbon emissions of software, there are several studies presenting calculation or measuring models for it [e.g. 21, 22, 23, 24, 25]. In this paper, we do not take a stand on which model should be used but only focus on the system of software-related emissions and the concept of stakeholder responsibilities for the emissions.

In their model, Hilty & Aebischer [14] also consider handprint effects, which are the benefits of ICT for other sectors, such as improving logistics through automation. This is also worth noticing when considering systemic issues, such as software-related energy consumption and $CO_2$ emissions, that there is another side of the coin existing.

*E. Shared responsibility*

Nollkaemper [26] states that "shared responsibility may be important for global governance in relation to such diverse areas as peacekeeping, climate change, migration, and conservation of natural resources". The term shared responsibility is used in multiple occasions e.g. areas concerning disaster or crisis management [27, 28], solidarity [29], or teaching [30]. According to Nollkaemper [26], shared responsibility refers to "situations where a multiplicity of actors contributes to a single harmful outcome, and legal responsibility for this harmful outcome is distributed among more than one of the contributing actors."

In previous research within the IT domain, shared responsibility has emerged as a significant concern in the realms of cloud services and security. Kartit et al. [31] introduced a model delineating the distribution of security responsibilities for various cloud services, namely IaaS, PaaS, and SaaS (see Fig. 3). Major cloud providers such as AWS[11], Google[12,] and Microsoft[13] also outline their perspectives on the division of cloud security responsibilities on their respective websites. Notably, these divisions align with Kartit et al.'s model presented in Figure 3. Since the enactment of the General Data Protection Regulation (GDPR)[14] in 2016, there has been a heightened focus on data security. This shift in attention has necessitated the precise definition of responsible stakeholders.

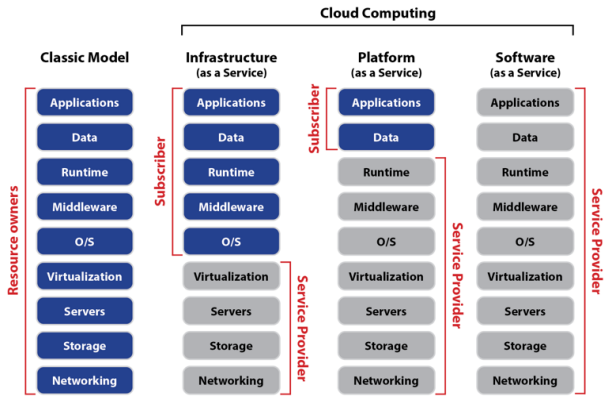

Fig. 3. Separation of responsibilities in the cloud according to Kartit et al. [31].

Additionally, AWS [32] introduces the concept of shared responsibility in cloud sustainability. According to AWS, the responsibility for sustainability of the cloud lies with AWS, while customers are responsible for sustainability in the cloud.

Partanen et al. [33] are first with their Shared Responsibility of Software Emissions (SRoSE) Framework to suggest shared responsibilities of software emissions (Fig. 4). The framework introduces three stakeholders, three sources, and five phases. Named stakeholders in the framework are the procurer, producer, and end user, and the named sources of the emissions are devices, networks, and servers. The framework has a timeline divided into five phases. These phases are procurement, development, deployment, usage, and end of lifecycle. The SRoSE Framework identifies direct effects and effects to be taken into account from the sources in different phases and suggests a responsible stakeholder in each phase [33].

## III. RESEARCH PROCESS

This research adopts Joseph A. Maxwell's model for qualitative research [34] in which research questions are not merely the starting point but are central to the research process (see Fig. 5). In this model, research questions play a pivotal role, influencing and being influenced by other components of the research framework, which are goals, conceptual framework, methods, and validity. Goals, methods, and validity are discussed further in this section while section I presents the research question and section II forms the conceptual framework.

The research question of this study presented in Section I emerged from insights gained during the activities in projects A and B. The projects included the following events during the years 2022-2023, providing knowledge and understanding, n being the number of individual participants.

- 32 interviews (n=48)
- 20 workshops (n=19)

The interviews were conducted as group or individual semi-structured interviews. Interviewees included procurer representatives (n=17), both public and private, and company representatives (n=29) including both software companies and IT consultancy organizations from SMEs to large companies. Participants from procurers included participants of legal (n=3), experts (n=2), middle management (n=9) and production (n=3). Participants from companies represented upper management (n=4), sustainability management (n=2), middle management (n=5), experts (n=1), and production (n=6).

Workshops included three types of internal project workshops done during projects A and B. Two sets with a total of 13 workshops were conducted in Project A and one set with seven workshops in Project B. Workshops were arranged for the development of two different artifacts, one in each project, and included experts from academy (n=9), non-profit organization (n=5), public organization (n=3) and private organization (n=2).

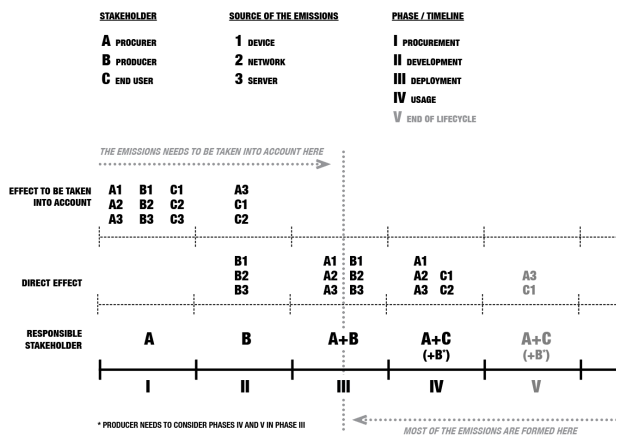

Fig. 4. Shared Responsibility of Software Emissions (SRoSE) framework [33].

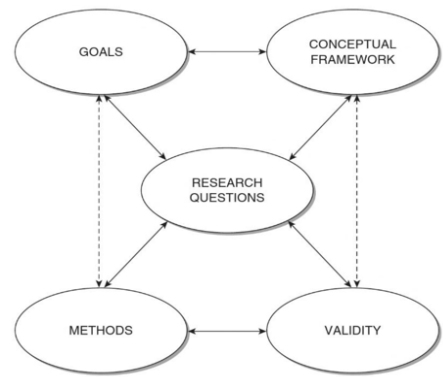

Fig. 5. Model for qualitative research design according to Maxwell [34].

---

[11] https://aws.amazon.com/compliance/shared-responsibility-model/
[12] https://cloud.google.com/architecture/framework/security/shared-responsibility-shared-fate
[13] https://learn.microsoft.com/en-us/azure/security/fundamentals/shared-responsibility
[14] https://eur-lex.europa.eu/legal-content/EN/TXT/HTML/?uri=CELEX:32016R0679

Theme raised from the interviews and workshops was the lack of responsibility for environmental impacts, but also the lack of possibilities to impact the decisions of other stakeholders. The goal of this research is to understand the complex system behind these themes.

In the research process, systems thinking was utilized. Systems thinking is an approach that focuses on understanding complex situations by examining wholes, patterns, and interrelationships, enabling the identification of root causes and innovative solutions and is applied e.g. in health research to help understand the complexity of health challenges, such as neglected tropical diseases. [35] According to Arnold and Wade [36], a system consists of elements with interconnections and purpose, and systems thinking itself is a system that builds up for the eight elements presented in Table II. This method fits well with sustainability and software since both are multidimensional, complex subjects having parts linking to each other. The utilization of the method in this study is presented in Table II.

In addition to the project interviews and workshops, the following individual actions have expanded but also validated the systems mapping during the end of the year 2023 and the year 2024.

- procurer interview (n=2)
- workshop with a software company representative (n=1)
- open discussion with academic peers (n=5)

Also, discussions at three different conferences and events arranged for doctoral students have influenced the systems thinking process. The contribution of all the events for each phase of the process is indicated in Table II with initials I, W, and D, where I = interviews, W = workshops, and D = discussions.

TABLE II. THE EIGHT ELEMENTS OF SYSTEMS THINKING [36] IN THE PROCESS OF THIS STUDY.

| Elements of systems thinking | Process phases in this study |
|---|---|
| 1) Recognizing interconnections | Identifying key connections through interviews and workshops (connections between stakeholders, connections between software and emission sources). [36] (I, W) |
| 2) Identifying and understanding feedback | Identifying and understanding the impact of cause-effect feedback loops (more software usage causes more emissions). [36] (I, W, D) |
| 3) Understanding system structure | Understanding the elements and interconnections in the system (understanding the big picture). [36] (I, W, D) |
| 4) Differentiating types of stocks, flows, variables | Defining and understanding stocks (energy required e.g. in data center), flows (amount of energy consumption) and variables (software usage affecting the energy consumption in data center). [36] (W, D) |
| 5) Identifying and understanding non-linear relationships | Identifying non-linear relationships (e.g. relationship between end-user and data center provider). [36] (I, W, D) |
| 6) Understanding dynamic behaviour | Understanding the impacts and effects in the system in different order of effects. [14] [36] (W, D) |
| 7) Reducing complexity by modelling systems conceptually | Intuitive [37] simplification by visualization, excluding the excess complexity (figures 8-11). [36] (W) |
| 8) Understanding systems at different scales | Understanding the levels in the systems (software level, stakeholder level), understanding the applicability in different contexts. [36] (D) |

IV. RESULTS

As a result of the research process presented in Table II, the mapping of software-related emissions, along with recommendations for the distribution of responsibilities, are presented in Figure 8. The following chapters provide a detailed examination of this mapping from multiple perspectives, supported by the visual representations in Figures 6–8.

A. Parts of the system

Parts of the system can be identified as stakeholders and artifacts. As presented in Section II the three main sources of emission are devices, data centers, and networks. These belong to artifacts. Partanen et al. [33] describe these also as sources in their SRoSE framework. However, the framework leaves these without a responsible stakeholder. To be able to draw the system mapping of the software-related emissions we name a responsible stakeholder to these artifacts as *the Provider* (Fig. 6). There exist providers for each of the three subjects – a device provider, a data center provider, and a network provider.

As *the Provider* artifacts (devices, data centers, networks) have their lifecycle, *the Software* also has its own, including different phases from development to usage and decommissioning [38, 39, 40]. Referring to Section II this can be seen as the actions and responsibilities of a customer. In our work, the role of a customer embodies both *the Procurer* and *the Producer* [33]. *Procurers* are the parties buying *the Software* and can be public or private organizations. Important is that the need for *the Software* Artifact comes from *the Procurer* (Fig. 7). It is worth mentioning that one might have both procurer and producer roles in subcontracting chains. With *the Producer,* we refer to all the parties involving the development of *the Software* artifact (Fig. 6). In practice, these are IT companies. There can be different parties as producers in different phases of the development process – e.g. one company is accountable for the design, one for the development, and one for the maintenance.

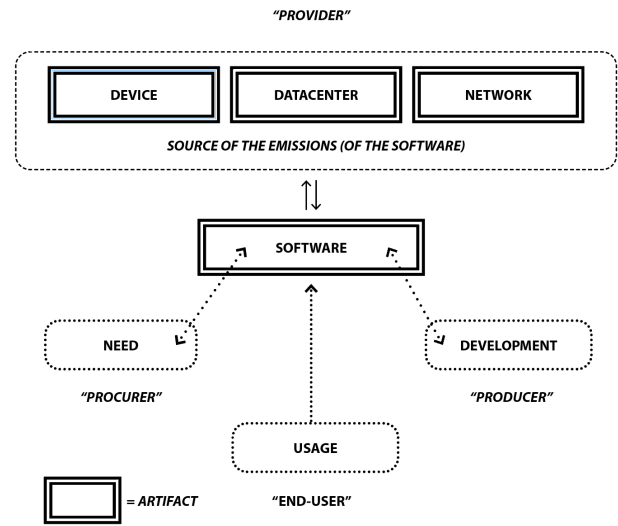

Fig. 6. Connections between parts of the system of software-related emissions.

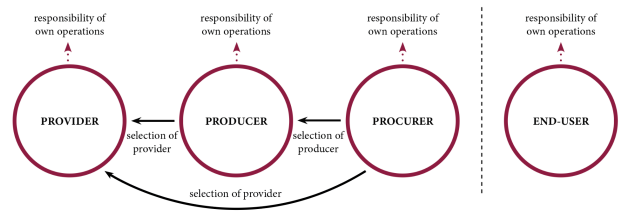

Fig. 7. Responsibilities between the four stakeholders.

*The End-user* is the party using the final product. The analysis has led to conclusions about the relationships and roles of the four stakeholders – *The Provider, the Producer, the Procurer,* and *the End-user. The End-user* stands apart from the three other stakeholders in that it is invariably an individual, and also the responsibility for emissions differs in nature from the other stakeholders, which operate at an organizational level. Additionally, regulations like CSRD do not extend to cover individuals. With positive reciprocal actions from individuals, end-users could initiate cooperation [41]. Likewise, end-users may find the product as an opportunity with features such as testimonials, ratings, reviews, or endorsements. Also, they can be committed with sign-up, survey, etc. in the product life-cycle that builds trust and consistency among them.

### B. Responsibilities

Recognizing the impracticality of collective responsibility [26], it becomes imperative to distribute this responsibility among the organizational stakeholders (*the Provider, the Producer, the Provider*). It is noticeable that every stakeholder in the equation needs to pay attention to not only external but also internal impacts. First, every stakeholder is responsible for their own operations and carbon footprint. Secondly, each stakeholder is responsible for selecting a suitable partner (see Fig. 7). This, however, requires transparency, traceability, reporting, and comparability. Every one of these needs elaborating, and the pressure coming from the EU regulation drives towards this.

### C. Mapping of the system of software-related emissions and responsibilities

The mapping of software-related emissions, responsibilities, and their interrelations is illustrated in Figure 8. The figure highlights various interconnections among stakeholders, as well as between stakeholders and artifacts, using distinct line styles to represent the nature of these relationships. First, each stakeholder is connected to others, either directly or indirectly, with the quality of these connections defined as *a choice*, reflecting their responsibilities of choosing a suitable partner or artifact. Secondly, stakeholders are accountable for their operations, as described in Section B with a second style of lines (arrows). Thirdly, the artifacts are also interconnected, directly or indirectly, with the nature of these connections characterized as *functionality* since to function, software needs an environment to function – a device. Nowadays, software most commonly also needs data transfer, which requires networks for the transfer and data centers for storing the data. To function, *the Software* needs *the Provider* artifacts (Fig 8). In turn, it can be thought that devices, data centers, and networks are needed because of software – if there were no software, there would not be the need for the infrastructure.

## V. DISCUSSION

Through the mapping of software-related emissions, we were able to give answers to the research question of what are the sources of software-related emissions, how are these sources interconnected, and what type of responsibilities the relevant stakeholders involved have in managing these emissions. Discussions with representatives from various perspectives in the field have provided valuable insights to affirm the mapping.

### A. The types of responsibilities

The responsibilities identified in the mapping can be divided into three themes – manufacturing and management of artifacts, organizations' own operations, and choice of artifact or partner. Responsibilities lying on the shoulders of the end-user are delimited to own operations and choices of the artifacts, basically devices and software. The nature of end-user responsibilities differs from the organizational responsibilities, as discussed in the paper, and therefore, discussion will be limited from an end-user perspective.

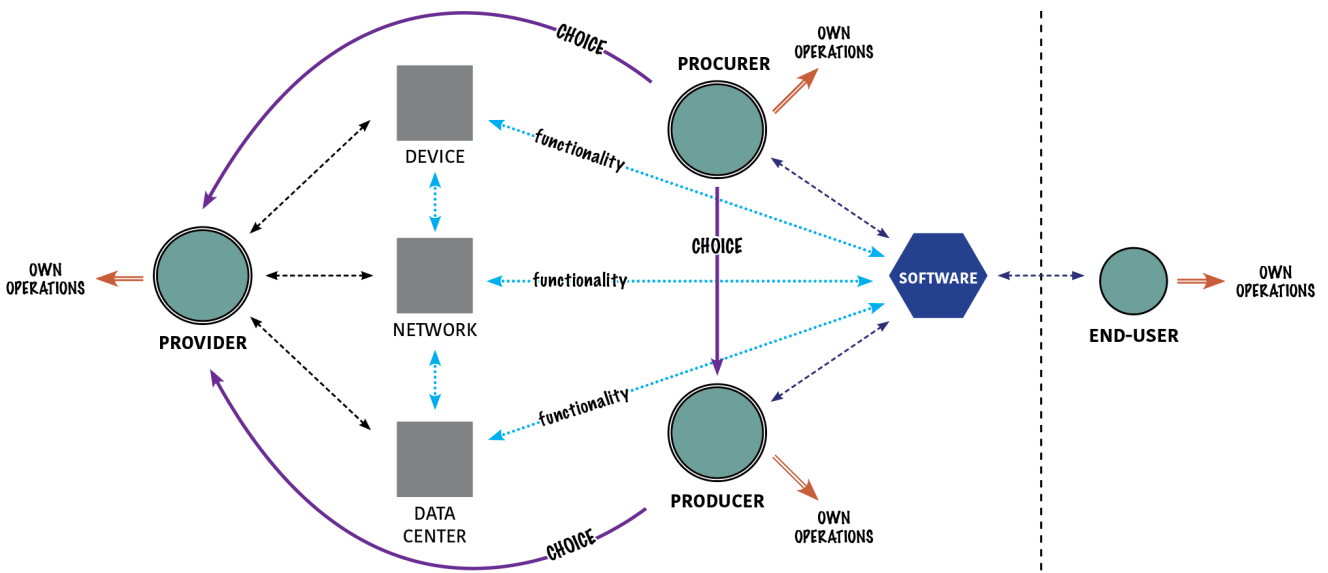

Fig. 8. Mapping of software-related emissions and responsible stakeholders within the system.

The provider of an artifact is responsible for the manufacturing of the artifact – the computer manufacturer bears responsibility for the lifecycle of the computer, including the materials used, the company processes, and the supply chain. The producer of software is responsible for implementing sustainable practices during software development, a conscious choice in itself. The procurer, too, holds responsibility for the choices made in selecting an appropriate partner. It can be argued that, in the end, the responsibility of sustainability lies in making sustainable choices. Ultimately, all related stakeholders bear accountability for their choices, whether internal or external.

Each stakeholder is responsible for the decisions made within the organization. For example, all stakeholders are accountable for decisions related to the devices they use. While they may not be responsible for the manufacturing of the devices, they do bear responsibility for the choices they make. Additionally, all stakeholders are accountable e.g. for the energy they choose to use and their organization policies.

From this, it can be concluded that while each stakeholder holds specific responsibilities, no single stakeholder is accountable for all aspects – the responsibility is shared. Awareness of the system of emissions and the distribution of responsibilities can foster a shift in mindset towards sustainability, ultimately encouraging more sustainable practices [42]. To advance this effort, greater attention and transparency are required, which will, in turn, improve traceability and accountability. Cooperating with stakeholders to create value remains a key challenge, but involving stakeholders and addressing their feedback and needs could promote the sustainable development of software services [43].

*B. 2-3-$5^2$ thinking*

It is discussed multiple times in the paper that one should also be aware of other aspects and dimensions of the impacts of ICT. For this, we suggest a memory rule of 2-3-$5^2$, where the number two stands for the two sides of the effects – the footprint and the handprint. The number three stands for the order of effects having direct, indirect, and systemic effects. [14] Finally, the number five stands for all listed environmental effects – climate change, land and sea usage, direct exploitation of organisms, pollution, and invasive alien species [7]. The second interpretation of the number five is five dimensions of sustainability – environmental, social, economic, technical, and individual [15, 16]. The memory rule can help recognize the linkages for software-related impacts.

*C. Future work*

While this study has provided significant insights into shared responsibilities of software-related emissions and energy consumption, several questions remain unanswered, offering opportunities for future research. The stakeholders considered in this study are highly simplified, whereas, in practice, the reality involves a far more complex and multifaceted network. Also, every source of the emission includes its components, which can be defined and studied deeper. Case studies would offer beneficial information to all of these questions.

Although the system mapping presented focuses primarily on carbon emissions, it has potential applications beyond this scope. For instance, it could be extended to encompass broader environmental dimensions, such as the overall environmental footprint. Additionally, the impacts could be examined from the perspective of other sustainability dimensions. Also, a broader mapping of each stakeholder's environmental responsibilities would benefit the practitioners.

VI. CONCLUSION

This study highlights the critical importance of addressing software-related emissions and the shared responsibilities among relevant stakeholders. Through the mapping of the system of software-related emissions, this research demonstrates that the emissions formed are not limited to a single source but stem from a system involving devices, data centers, networks, and the broader ICT infrastructure. This research also introduces relationships and responsibilities across the identified four key stakeholders: providers, producers, procurers, and end-users. Each stakeholder must not only manage their direct emissions but also make informed decisions when selecting partners within the value chain. The findings emphasize that while end-users play a pivotal role, the majority of responsibilities lie with organizations. Sustainability reporting, driven by evolving EU regulations such as the CSRD, is essential for ensuring accountability.

ACKNOWLEDGMENT

This research has been financially supported by the PHP Säätiö.

REFERENCES

[1] Ranasinghe, H.: Carbon Net-Zero by 2050: Benefits, Challenges and Way Forward. Journal of Tropical Forestry and Environment, 12(01) (2022).

[2] Huttunen, R., Kuuva, P., Kinnunen, M., Lemström, B., Hirvonen, P.: Carbon neutral Finland 2035 – national climate and energy strategy. Publications of the Ministry of Economic Affairs and Employment 2022:55 (2022).

[3] Freitag, C., Berners-Lee, M., Widdicks, K., Knowles, B., Blair, G. S., Friday, A.: The real climate and transformative impact of ICT: A critique of estimates, trends, and regulations (2021).

[4] Issa, T., Isaias, P.: Sustainable Design. Springer-Verlag London Ltd. (2022).

[5] Global e-Sustainability Initiative: #SMARTer2030. ICT Solutions for 21st Century Challenges. https://smarter2030.gesi.org/downloads/Full_report.pdf, last accessed 2024/09/08.

[6] Ojala, T., Oksanen, P.: Climate and Environmental Strategy for the ICT Sector. Publications of the Ministry of Transport and Communications 2021:6 (2021).

[7] IPBES: Summary for policymakers of the global assessment report on biodiversity and ecosystem services of the Intergovernmental Science-Policy Platform on Biodiversity and Ecosystem Services. IPBES secretariat, Bonn, Germany. 56 pages (2019)

[8] Sipilä, A., Partanen, L., Porras, J.: Carbon footprint calculations for a software company – adapting GHG Protocol scopes 1, 2 and 3 to the software industry. Software Business: 14th International Conference, ICSOB 2023, Lahti, Finland, November 27-29, 2023, Proceedings. Springer Nature (2023).

[9] Brundtland, G.H.: Our Common Future: Report of the World Commission on Environment and Development. Technical Report A/42/427, United Nations, (1987).

[10] James, P.: Urban Sustainability in Theory and Practice: Circles of sustainability. Routledge; 283 p. (2014).

[11] Richardson, K., Steffen, W., Lucht, W., Bendtsen, J., Cornell, S.E., Donges, J.F., Drüke, M., Fetzer, I., Bala, G., von Bloh, W., Feulner, G., Fiedler, S., Gerten, D., Gleeson, T., Hofmann, M., Huiskamp, W., Kummu, M., Mohan, C., Nogués-Bravo, D., Petri, S., Porkka, M., Rahmstorf, S., Schaphoff, S., Thonicke, K., Tobian, A., Virkki, V., Weber, L. & Rockström, J.: Earth beyond six of nine planetary boundaries. Science Advances 9, 37 (2023).

[12] Dasgupta, P.: The Economics of Biodiversity: The Dasgupta Review. Abridged Version. (London: HM Treasury) (2021).

[13] Coroama V.C., Hilty L.M.: Energy consumed vs. Energy saved by ICT-a closer look EnviroInfo (2), pp. 347-355 (2009).

[14] Hilty L.M., Aebischer B.: ICT for sustainability: An emerging research field ICT Innovations for Sustainability. Springer (2015), pp. 3-36 (2015).

[15] Penzenstadler, B., Raturi, A., Richardson, D. & Tomlinson, B.: Safety, security, now sustainability: The nonfunctional requirement for the 21st century. IEEE software, 31(3):40–47 (2014).

[16] Becker C., Chitchyan R., Duboc L., Easterbrook S., Penzenstadler B., Seyff N., Venters C.C.: Sustainability design and software: The karlskrona manifesto. IEEE (2015), pp. 467-476 (2015).

[17] Duboc L., Penzenstadler B., Porras J., Kocak S. A., Betz S., Chitchyan R., Leifler O., Seyff N. & Venters C. C.: Requirements engineering for sustainability: An awareness framework for designing software systems for a better tomorrow. Requirements Engineering 25, 4 (2020), 469–492 (2020).

[18] Salonen, L.: Finding Ecologically Sustainable Design Principles for Mobile Devices – Towards a Heuristic Model from a List of Good Design Practices (2021)

[19] The World Bank and ITU: Measuring the Emissions & Energy Footprint of the ICT Sector: Implications for Climate Action. Washington, D.C. and Geneva (2024).

[20] Green Software Practitioner. https://learn.greensoftware.foundation/measurement/, last accessed 2024/09/06.

[21] Taina, J.: How Green Is Your Software? Lecture Notes in Business Information Processing. 51. pp. 151-162 (2010).

[22] Simon, T., Rust, P., Rouvoy, R., Penhoat, J.: Uncovering the Environmental Impact of Software Life Cycle. International Conference on Information and Communications Technology for Sustainability, Rennes, France (2023).

[23] Kern, E., Dick, M. Naumann, S., Hiller, T.: Impacts of software and its engineering on the carbon footprint of ICT. . Environmental Impact Assessment Review. 52 (2014).

[24] Gupta, U., Kim, Y. G., Lee, S., Tse, J., Lee, H-H S., Wei, G-Y, Brooks, D., Wu, C-J.: Chasing Carbon: The Elusive Environmental Footprint of Computing. In: CoRRabs/2011.02839. (2020).

[25] Naumann S, Dick M, Kern E, Johann T.: The GREENSOFT Model: a reference model for green and sustainable software and its engineering. Sustainable Computing: Informatics and Systems 2011;1(4). pp. 294–304. (2011).

[26] Nollkaemper, A.: The duality of shared responsibility. Contemporary Politics, 24:5, 524-544 (2018).

[27] McLennan, B., & Eburn, M.: Exposing hidden-value trade-offs: Sharing wildfire management responsibility between government and citizens. International Journal of Wildland Fire, 24, 162–169 (2015).

[28] Lukasiewicz, A., Dovers, S. & Eburn, M.: Shared responsibility: the who, what and how. Environmental Hazards. 16. 1-23 (2017).

[29] ACVZ: Sharing responsibility - a proposal for a European asylum system based on solidarity. Publication of the ACVZ, The Hague, 2015. Advisory report reference: 43•2015, (2015).

[30] Gourvennec, A.F., Solheim, O.J., Foldnes, N., Uppstad, P. & McTigue, E.: Shared responsibility between teachers predicts student achievement: A mixed methods study in Norwegian co-taught literacy classes. J Educ Change (2022).

[31] Kartit, Z., Idrissi, H., El Marraki, M., & Ali, K.: Network Issues in cloud computing and countermeasures. JNS4 Tetouan (2014).

[32] Philipp, K., Yunus, A., Antoniou, O. & Tahtasiz, C.: Optimizing your AWS Infrastructure for Sustainability, Part I: Compute. 2021. https://aws.amazon.com/blogs/architecture/optimizing-your-aws-infrastructure-for-sustainability-part-i-compute/, last accessed 2024/08/27.

[33] Partanen, L., Sipilä, A., Porras, J.: Energy Consumption and CO2 Emissions of a Software – Who is Responsible? ICSOB '23: 14th International Conference on Software Business, November 27–29, 2023, Lahti, Finland. CEUR-WS.org (2023).

[34] Maxwell, J. A.: Qualitative Research Design: an Interactive Approach. Thousand Oaks, California. SAGE Publications (2013).

[35] Glenn J, Kamara K, Umar ZA, Chahine T, Daulaire N, Bossert T. Applied systems thinking: a viable approach to identify leverage points for accelerating progress towards ending neglected tropical diseases. Health Res Policy Syst. 2020 Jun 3;18(1):56 (2020).

[36] Arnold, R. D. & Wade, J. P.: A Definition of Systems Thinking: A Systems Approach. Procedia Computer Science, Volume 44, p. 669-678 (2015).

[37] Raami, A.: Experiences on Developing Intuitive Thinking among University-level Teachers. EKSIG 2013 Conference, DRS Design Research Society's Special Interest Groups on Experiential Knowledge "Knowing Inside Out - experiential knowledge, expertise and connoisseurship" UK, Loughborough University (2013).

[38] Kivimäki, L., Partanen, L., Porras, J., Tarkkanen, K., Tuikka, A-M, Mäkelä J-M & Mäkilä, T.: Building Up Green Software Life Cycle Model (forthcoming).

[39] Bourque, P. & Fairley, R.E.: Guide to the Software Engineering Body of Knowledge - SWEBOK V3.0 (2014).

[40] ISO Central Secretary. Standard ISO/IEC/IEEE 12207:2017(E). Systems and software engineering – Softwarelife cycle processes. International Organization for Standardization, Geneva, CH (2017).

[41] Umetani R., Yamamoto H., Goto A., Okada I. & Akiyama E.: Individuals reciprocate negative actions revealing negative upstream reciprocity. PLoS One. 2023 Jul 5;18(7):e0288019 (2023).

[42] Porras, J., Abdullai, L., Partanen, L. & Sipilä, A.: Towards an Ecosystem Model in Enabling Sustainability Transition: The Case of Finnish Government Sustainability Strategy for the ICT Sector (forthcoming).

[43] Samant, S., Sangle, S. & Daulatkar, S.: Co-Creation with Stakeholders: The Key to Enhancing Sustainable Value. International Journal of Social Ecology and Sustainable Development (IJSESD), IGI Global, vol. 7(3), pages 34-46, July (2016).